# An Efficient Playout Smoothing Mechanism for Layered Streaming in P2P Networks

Abbas Bradai[1], Ubaid Abbasi[1], Raul Landa[2], and Toufik Ahmed[1]

[1] CNRS-LaBRI, University of Bordeaux-1, France
351, Cours de la Libération
Talence Cedex, France
{bradai, abbasi, tad} @labri.fr

[2] Department of Electronic and Electrical Engineering, University College London,
Torrington Place, London WC1E 7JE,
United Kingdom
rlanda@ee.ucl.ac.uk

*Abstract*- Layered video streaming in peer-to-peer (P2P) networks has drawn great interest, since it can not only accommodate large numbers of users, but also handle peer heterogeneity. However, there's still a lack of comprehensive studies on chunk scheduling for the smooth playout of layered streams in P2P networks. In these situations, a playout smoothing mechanism can be used to ensure the uniform delivery of the layered stream. This can be achieved by reducing the quality changes that the stream undergoes when adapting to changing network conditions. This paper complements previous efforts in throughput maximization and delay minimization for P2P streaming by considering the consequences of playout smoothing on the scheduling mechanisms for stream layer acquisition.

The two main problems to be considered when designing a playout smoothing mechanism for P2P streaming are the fluctuation in available bandwidth between peers and the unreliability of user-contributed resources – particularly peer churn. Since the consideration of these two factors in the selection and scheduling of stream layers is crucial to maintain smooth stream playout, the main objective of our smoothing mechanism becomes the determination of how many layers to request from which peers, and in which order.

In this work, we propose a playout smoothing mechanism for layered P2P streaming. The proposed mechanism relies on a novel scheduling algorithm that enables each peer to select appropriate stream layers, along with appropriate peers to provide them.

In addition to playout smoothing, the presented mechanism also makes efficient use of network resources and provides high system throughput. An evaluation of the performance of the mechanism demonstrates that the proposed mechanism provides a significant improvement in the received video quality in terms of lowering the number of layer changes and useless chunks while improving bandwidth utilization.

*Index Terms*- Peer-to-Peer (P2P) Network, layered streaming, smoothing, scheduling, quality of experience (QoE)

## I. INTRODUCTION

Over the past few years, layered video streaming has emerged as a promising approach for distribution of multimedia content in P2P networks. The usefulness of layered video coding arises from its ability to support large number of users, while simultaneously handling client heterogeneity in terms of download bandwidth, terminal capabilities and user preferences. With layered video coding, the original video is partitioned into multiple layers, which are then transmitted independently. This allows peers with high capacity to receive all layers of the video and enjoy maximum quality, while peers with lower capacity receive a subset of layers and experience reduced quality.

In order to support layered video streaming in P2P networks, three essential components need to be considered: *overlay construction, content delivery* and *content adaptation.* The first two are basic building blocks in most peer-to-peer systems, and as such have received ample attention from the research community. However, in order to tackle heterogeneity in terms of terminal capabilities, network conditions and user preferences*, content adaptation* is rapidly becoming a core component for such systems. Traditionally, the overlay construction component deals with the selection of appropriate neighbors for the retrieval of content*,* and the content delivery component is responsible for the requesting and transport of content chunks from the chosen overlay neighbors. In this work we present innovations on both the content adaptation component and its relationship with overlay construction and content delivery. To this end, we consider the appropriate layer selection for layered streaming systems and how to request the chunks of those layers from the overlay neighbors.

The design of algorithms that optimally perform these tasks is non-trivial, especially when uniform, high quality playback is desired. Of the many engineering challenges posed by such design, one of the most important is the fluctuation in available bandwidth between peers. On one hand, the delivery of maximal video quality to the user provides a rationale for the algorithm to aggressively select higher quality layers when sufficient bandwidth becomes available. On the other hand, if this bandwidth is only available for a brief period, the algorithm will soon be forced to fall back to selecting lower quality layers, leading to an undesirable fluctuation in user QoE (Quality of Experience). Therefore, a playout smoothing mechanism must balance the aggressiveness with which it uses bandwidth when it becomes available, and the conservativeness with which it maintains a stable user QoE.

In non-layered streaming, there's almost no difference between high delivery ratio and high throughput, because there is no inter-layer dependency. This is not the case in layered streaming. For instance, under certain bandwidth conditions, the following two scenarios can result in the same throughput, but not the same quality: (1) to select many layers, experiencing low delivery ratio for each one of them, and (2) to select fewer layers, but experiencing a higher delivery ratio for each one of them. It's obvious that although the same throughput is achieved by both methods, the former one is undesirable due to its low delivery ratio for each layer. This problem is further compounded because of inter-layer dependencies: Since enhancement layers are only useful if lower layers are present, losses at the lower layers can severely degrade system performance and lead to wasted system resources if enhancement layers are selected for which the corresponding lower layers are missing. We will refer as *useless chunks* to these upper layer chunks that cannot be correctly decoded because some lower layer chunks are missing.

In order to ensure effective bandwidth utilization, the scheduling mechanism should also take into consideration the availability of chunks in the overlay neighbors as well as the link capacity between the receiver peer and its neighbors. The challenge is, then, to find an effective scheduling algorithm that provides stable QoE while fully taking advantage of the available network capacity. In this work, we present a mechanism which achieves the following design objectives:

*Smoothing:* To design a quality adaptation mechanism with the ability to control the level of smoothness. Having such a tuning capability, one can tune the quality adaptation mechanism for layered video encoding to minimize the effect on the perceived quality by adding and dropping layers and switching gracefully from one quality level to another.

*Efficiency:* To design a better scheduling mechanism to ensure the best use of the available download bandwidth in different links taking into consideration the chunks availability in the neighborhood, the urgency of chunks, and the dependencies between layers. The main goal of our scheduling algorithm is to minimize the useless chunk ratio by introducing a priority mechanism for proper delivery of different layers. Thus, the sequence in which chunks of different layers are requested is a critical issue and should be tackled by an efficient scheduling mechanism.

In our work we focus on the bandwidth variation problem and its impact on the quality level fluctuation in layered P2P streaming. We posit that, in addition to content availability on neighbors, the maximum achievable quality level depends critically on the available download bandwidth on the receiver peer. Hence, we propose an algorithm to select the maximum quality level based on the available bandwidth. This is by efficiently requesting chunks from neighbors while at the same time maintaining a satisfactory level of video smoothness.

We continue with our presentation as follows. First, we present a mechanism for smooth selection of appropriate layers considering the available bandwidth fluctuations in the network. Then, we propose a chunk prioritization strategy that considers the urgency of chunks and the dependencies between the layers to which they belong. We then model chunk scheduling as an assignment problem, and propose an algorithm for its solution that takes into account both the available bandwidth and chunk availability for each peer. The rest of this paper is structured as follows: section II reviews the related work in the area of layered streaming in P2P networks. In order to understand the problem of smoothing and scheduling in layered streaming, we describe problem statement in section III. In section IV, we present the proposed mechanism of playout smoothing for layered streaming in P2P networks. Section V provides the detailed performance analysis of our proposed mechanism. At last, section VI briefly concludes the work.

II. RELATED WORK

There have been numerous efforts in the design and evaluation of layered video streaming systems in the last decade. In addition to its promise in handling client heterogeneity, layered video has recently received attention as an application layer solution to the limited deployment of IP-multicast. We now present some of these works, and contrast them with the one presented in this paper.

To handle changing networks dynamics, several layered streaming systems have been proposed [1][2][2][4][5][5]. For instance, PALS [4], proposed by Rejaie et al. focused on another dimension of the layered P2P streaming problem. It is a receiver driven P2P video streaming system with quality-adaptive playback of layered video. The system provides an adaptive streaming mechanism from multiple senders to a single receiver. It enables a receiver peer to orchestrate quality adaptive streaming of layered encoded video stream from multiple congestion controlled senders, and is able to support a spectrum of non-interactive streaming applications. However, PALS didn't consider the smoothing problem due to bandwidth variation, nor the assignment of chunk requests to appropriate peers. Another example is [5] that focused on combining the benefits of network coding with layered streaming to mitigate the inherent challenges in unstructured P2P systems. The work focuses on the average quality satisfaction of the peer, but does not consider the degradation in user's quality of experience (QoE) due to variation in quality levels.

Recently, Fernandes et al. [7] proposed a mechanism for scalable streaming of stored video over the networks with explicit feedback notification. The mechanism considers the unpredictability of the available rate in order to determine the output sending rate target. The authors discussed the possibility of smoothing the information received from the transport layer before making any decision concerning the sending rate. The authors provide a solution to the problem of accommodating the mismatch between the available bandwidth variability and the encoded video rate variability.

Another contribution to the area is [8], in which the authors proposed a taxation-based P2P layered streaming design including layer subscription strategy and mesh topology adaptation. The taxation mechanism is devised to strike the right balance between social welfare and that of individual peers. Although the work considers the strategy for layer selection, it mainly focuses on fairness in P2P systems.

In [9], Nguyen et al. demonstrate the importance of neighbor selection in layered streaming and identify the unique challenges of neighbor selection for system performance. In addition, the authors propose a new neighbor selection technique that can offer good performance and scalability under network fluctuations. The core of the technique is a preemption rule that biases neighbor selection policies by taking into account peer capacity. The work focuses on achieving high quality by providing each peer with a set of neighbors having higher perceived quality. Our work moves a step further by not only selecting the appropriate layers according to available local resources to ensure smoothing, but also by ensuring the effective utilization of the available overlay capacity.

With regards to chunk scheduling in P2P networks, many works are based on empirical studies for specific policies and heuristics. Examples of this include a pure random strategy [9], Local Rarest First (LRF) [11] and Round Robin (RR)[12]. Apart from empirical studies, some works use queuing models for scheduling [13]. The algorithm proposed in [14] minimizes the base layer losses, but it assumes equal rates for the base and enhancement layers. This model of video is rather ideal and can be approximated only by fine grained scalability (FGS). Furthermore, a few theoretical studies tackle the optimal stream scheduling. Most of these works are under restrictive hypothesis or computationally expensive. In [15] a scheduler has been proposed to maximize the video quality by prioritizing the most important chunks. This strategy is particularly suited for push-based, tree-structured overlays. The scheduling mechanism proposed in LayerP2P [16] is able to save base layer losses to the detriment of the enhancement layers. Authors propose to categorize chunks request into two types: regular requests and probing requests. The regular request concerns the requests of layers lower than or equal to a threshold $l_n$, which are firstly assigned to different suppliers based on random scheduling algorithm (that authors believe it achieves a high system throughput) without any prioritization among different layers. Secondly, the probing requests (layer greater than $l_n$) are sent to the suppliers layer by layer, in ascending manner. The quality threshold $l_n$ and the maximum quality level to be requested are decided based of the available download capacity on the receiver peer, by consequence it follows the bandwidth fluctuation without any smoothing mechanism.

The authors in [17] propose an optimal scheduling strategy to minimize the overall video distortion, but the approach is strongly related to the Multiple Description (MD) coding, which is less efficient compared with layered coding [17]. Zhang et al. [19] have discussed the scheduling problem in data-driven streaming systems. They define a utility for each chunk as a function of its *rarity*, which is the number of potential senders of this chunk, and its *urgency*, which is the time difference between the current time and the deadline of this chunk. They then use this model to transform the chunk scheduling problem into a min-cost flow problem. This algorithm, however, is computationally expensive and may not be feasible for live video streaming systems subject to strict deadlines on computationally-constrained devices.

Szkaliczki et al. [20] also address the chunk selection problem in streaming layered video content over peer-to-peer networks. The authors present a number of theoretical solutions to maximize the utility function of chunks that exist in the literature. However, their proposed solutions rely on the definition of chunk utility functions whose objective definition may be difficult in real-life scenarios.

## III. PROBLEM STATEMENT

One of the problems in assessing the performance of a video delivery scheme is the lack of a good metric that captures the user's perception of video quality. It is generally observed that it is visually more pleasing to watch a video with consistent, lower quality than one with higher but varying quality [21]. However, reducing the quality to a bare minimum by following a strictly conservative approach is undesirable, as it fails to adequately take advantage of available overlay resources. The objective of the layer selection mechanism presented in next sections is to optimize the perceived video quality, while at the same time ensuring the smooth delivery of the layered stream. To explain our smoothness criterion, we direct the reader to Figure 1, which exemplifies two possible approaches to stream smoothing for a given available bandwidth profile.

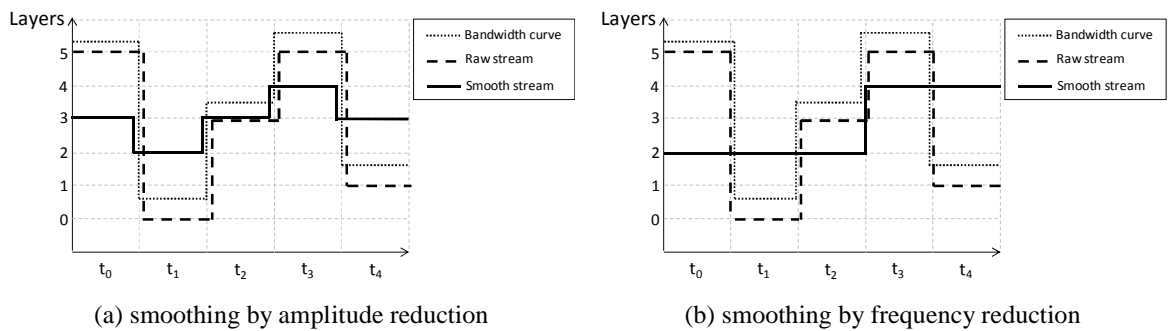

(a) smoothing by amplitude reduction     (b) smoothing by frequency reduction

Figure 1: Variation in quality level

In both Figure 1(a) and Figure 1(b), raw stream attempts to precisely track the changing available bandwidth. As a result, the QoE of the user may be severely degraded, especially when there is a drop from a high quality level to a much lower one (as in the case of time slot $t_1$ and $t_4$). In comparison, the smoothed stream depicted in Figure 1(a) reduces the size of the jump from higher to lower quality level. The objective here is to ensure a gradual change in quality levels, rather than subjecting the user to widely varying QoE. This technique is referred to as amplitude reduction. An alternative technique is shown in Figure 1(b), where smoothing focuses on reducing the number of quality changes from 4 in Figure 1(a) to 1 in Figure 1(b). This technique, referred to as *frequency reduction*, aims to reduce the number of changes in quality level due to variations in available bandwidth.

The playout smoothing mechanism should take into consideration two additional factors. Firstly, the smoothing mechanism should neither be too conservative (sacrificing higher quality to achieve long term smoothness) nor too aggressive (sacrificing better smoothness to better take advantage of short-term available bandwidth). Secondly, the smoothing mechanism should also take into the consideration the extra delay for the user that may experience as a side-effect of the smoothing algorithm. This extra-delay may adversely affect the liveness of the stream, thus making it unsuitable for live streaming applications. Thus, a playback smoothing mechanism should apply both amplitude and frequency reduction to achieve a good tradeoff between user QoE and bandwidth efficiency while incurring low processing delay. Once the playout smoothing mechanism has selected a target quality level, the next step for the algorithm is to decide the order in which the chunks of the selected layers are requested, and from which neighbor peers. This must be done in such a way that all higher layer chunks available in the decoding buffer can be decoded before their playback deadline expires. If, for any reason a higher layer chunk is acquired and is not decoded on time, resources have been wasted on it, and it is considered a *useless chunk*. In addition, the playout smoothing mechanism must also utilize available system resources efficiently.

To better explain the problem of scheduling, we assume a mesh-based pull approach in which the receiver side buffer is organized into a sliding window (Figure 2) containing chunks of different layers. The chunks beyond the playhead position are denoted as the *exchanging window*; only these chunks are requested if they have not been received yet (the chunks whose deadline has passed will not be requested). Each peer periodically announces the chunks that it holds to all its neighbors by sending a *buffer map* (Figure 3), a bit vector in which each bit represents the availability of a chunk in the sliding window. Periodically, each peer sends requests to its neighbors for the missed chunks in its exchanging window. As long as its request remains in the exchanging window, chunks are re-requested if not received.

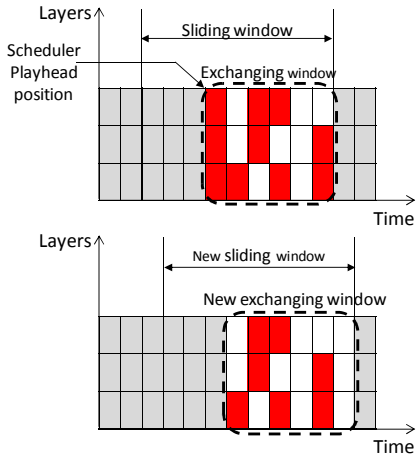
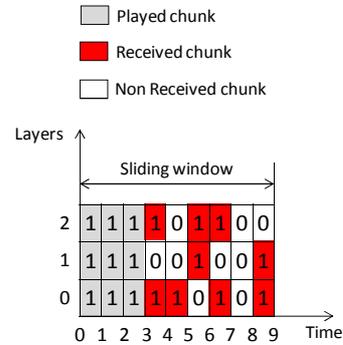

Figure 2: Sliding window mechanism

Figure 3: Buffer map structure

Of course, upper layer chunks received without the corresponding lower layer chunks are not decodable (and are considered useless, as described earlier). Thus, the chunks having time stamp $T = 5$ in Figure 3 are not played, because the base layer was not received.

In order to increase the throughput of the system, our approach aims to take full advantage of the download bandwidth of peers by maximizing the number of chunks that are requested within each scheduling period. Figure 4 illustrates an example of the optimal scheduling problem in terms of bandwidth utilization. For simplicity, in this example we consider a single-layer stream. Peer 1 is the receiving node, and it requests missing chunks from its neighbor peers 2, 3 and 4. Each neighbor advertises the chunks that it holds using a buffer-map. The numbers on the arcs denote the units of bandwidth that the neighbor peer is willing to provide to the receiver node (peer 1) in terms of chunks per unit time. An optimal scheduling scheme for this example is represented in Figure 5, where rows represent the peers and the columns represent the chunks. Chunk 1 is requested from peer 4, chunks 2 and 3 from peer 2, and chunks 4 and 5 from peer 3. This strategy takes full advantage of the available bandwidth of the network. In Figure 6, we represent the result of Round Robin scheduling strategy [12] applied to the same example. In this case, only 4 chunks out of the total of 5 can be requested in a single time unit.

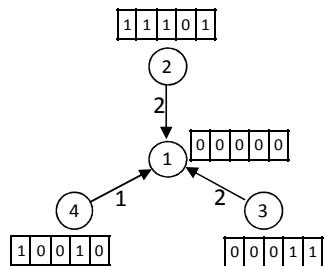

Figure 4: Example of the optimal chunk scheduling problem

| Chunk \ Node | 1 | 2 | 3 | 4 | 5 |
|---|---|---|---|---|---|
| 2 | 1 | 1 | 1 | 0 | 1 |
| 3 | 0 | 0 | 0 | 1 | 1 |
| 4 | 1 | 0 | 0 | 1 | 0 |

| Chunk \ Node | 1 | 2 | 3 | 4 | 5 |
|---|---|---|---|---|---|
| 2 | 1 | 1 | 1 | 0 | 1 |
| 3 | 0 | 0 | 0 | 1 | 1 |
| 4 | 1 | 0 | 0 | 1 | 0 |

Figure 5: Optimal Chunk scheduling example     Figure 6: Round robin scheduling example

The basic idea of our proposed mechanism is to select the appropriate enhancement layers based on the current quality level and an estimation of the available bandwidth for next time period. In addition, for each chunk belonging to a selected layer, we will define a *priority* on the basis of its playback deadline and its dependencies with other layers. This priority will then be used to guide chunk scheduling by requesting those chunks with higher priority first.

## IV. SCHEDULING FOR SMOOTH LAYERED STREAMING

In this section, we present the core concepts behind our proposed mechanism. We assume a chunk-based, mesh-based pull approach in which chunks are the basic unit of data exchange in the network; each chunk carries information for a given video segment at a given layer. The receiver peer requests content from its neighbors according to the basic architecture depicted in Figure 7. As shown, our proposed scheduling mechanism can be decomposed into two functions: *smoothing* and *scheduling*. First, the smoothing function defines the layers to be requested, taking into account the estimated available bandwidth at the receiver peer. In practice, this function operates over two distinct time horizons. The first one, which we will call the *initial* quality smoothing function, is invoked only once at the beginning of the session, and is responsible for the definition of the complete set of layers that the algorithm will consider at execution time. The second time horizon, which we will call the *runtime* quality smoothing function, is used during the execution of the algorithm to select, according to measured bandwidth variations, between the set of layers defined at startup.

Once the selection of appropriate layers for the next time period is performed, the scheduling function is then responsible for requesting the necessary chunks to achieve the selected quality level. To this end, this function assigns chunks with priorities according to their playback deadline and layer dependency. The output of this module is a chunk to peer assignment matrix as shown in Figure 7. The following sub-sections describe both modules in greater detail.

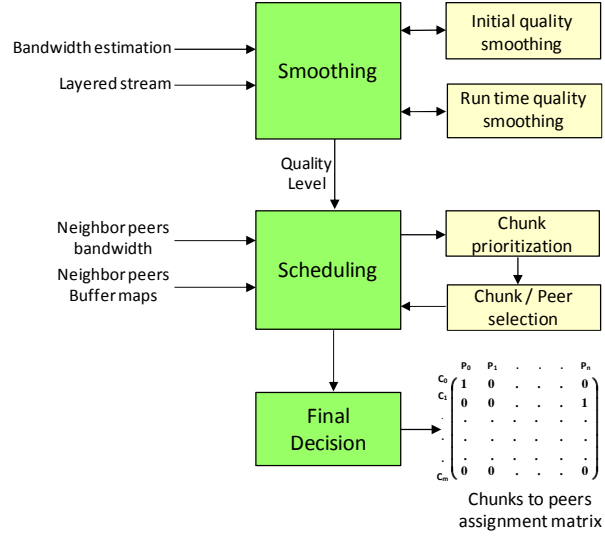

Figure 7: Workflow of proposed playout smoothing mechanism

## A. The Smoothing Function

The core objective of the smoothing function is to select which layers will be requested during next time period in order to simultaneously reduce the quality variability for the layered stream (to increase overall QoE) while increasing the overall stream throughput (both to increase overall quality and better make use of available network bandwidth). We will achieve this by first deriving a *tradeoff* quality level that reduces variability while increasing overall quality, subject to the constraint that the bandwidth required for achieving this quality level should not exceed the available bandwidth at the receiver peer.

We consider a discrete-time model, where each time slot represents an arbitrary number of video chunks. Let $L^t$ represent the selected tradeoff quality level at time *t*, and $S_W$ the smoothing window size.

In order to provide the smoothing algorithm with relevant information regarding video chunks, we divide the receiver side exchanging window into the following three different intervals (see Figure 8).

*Playing Buffer:* It contains a number of chunks ready for playing. The player makes decisions based on layer dependency on the basis of available chunks in this buffer.

*Urgent Buffer:* It contains a number of chunks that should be requested urgently from overlay neighbor peers. This buffer contains the smoothing window of length $S_W$, used by our proposed algorithm to define the tradeoff quality level $L^t$ to be used as input for the scheduling function. Only those chunks belonging to layers necessary to achieve $L^t$ are requested (i.e. $L^k \ \forall k \leq t$). The length of smoothing window introduces an extra lag time prior to decoding that should be taken into consideration for live streaming applications.

*Prefetching Buffer:* It contains a number of possibly useful chunks which can be prefetched in future for decoding the content.

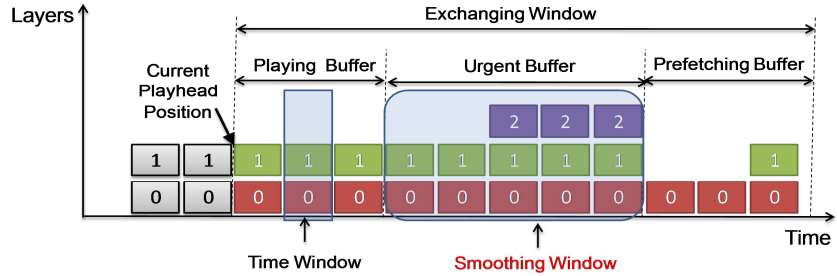
Figure 8: High level overview of sliding window at receiver peer

As stated previously, the smoothing function operates over two disjoint time horizons: Initial startup (the initial quality smoothing function) and real-time adaptation (the *runtime* quality smoothing function). The *startup* function is invoked only once, and it is responsible for selection of appropriate layers in the beginning of the session. It is designed in such a way that each peer can determine the highest quality level it can achieve before starting to play the layered video stream. The purpose of this function is to initialize the quality smoothing function with important information regarding the user context. The important considerations for this determination are user preferences, terminal capabilities, link capacity and video decoder processing power. Using different types of user metadata, we can filter out those layers that are not compatible with user request. The provisioning of metadata is out of the scope of this paper and not considered in this work.

Of course, due to changing network conditions, the maximum capability information obtained as a result of this initial layer selection process is not sufficient for most practical scenarios. Hence, the runtime quality smoothing function dynamically adjusts layer selection to implement both amplitude and frequency reduction according to variations in available network bandwidth. To implement amplitude reduction, first we estimate the download bandwidth of the receiving peer, and then request the missing chunks according to their priorities as described in next section. For the bandwidth estimation, we aim to utilize an Auto Regressive Integrated Moving Average (ARIMA) time series model that, given information up to time *t-1,* provides a forecast for time *t*. The classical approach by Box and Jenkins for modeling and forecasting ARIMA time series is performed in three steps [22]. First, model identification is used to estimate a model structure by using autocorrelation (ACF) and partial-autocorrelation (PACF) functions to expose dependencies among data. In this case, the major task is to transform a non-stationary series into a stationary one. Once model identification is done, a parameter estimation method is used to fit the identified model to the observed data. This is done by determining the coefficients of the linear model. The last step is then the prediction of future values. For simplicity in our evaluation, we used the bandwidth samples from the last smoothing window to estimate the bandwidth in the next time period. These available bandwidth values change as the sliding window proceeds.

The estimated bandwidth allows us to select the appropriate quality levels for next time period taking into consideration the amplitude variation. Conversely, to implement frequency reduction, we calculate the quality level that can be sustained for previous smoothing window, and use it to select the quality level for the next one.

We now present our algorithms for both amplitude and frequency reduction, and then a hybrid approach that provides the benefits of both.

*i. Amplitude Reduction*

The objective of amplitude reduction is, essentially, the reduction in the size of jumps between quality levels. If we assume that quality level jumps are a random variable, this is essentially equivalent to reducing their dispersion around the mean. Therefore, when measuring jump size, we consider the mean quality level $\bar{L}$ achieved during the previous smoothing window. In particular, $\bar{L}$ is defined as:

$$\bar{L} = \frac{1}{S_W}\sum_{j=t_c-s_w}^{j=t_c} L^j \qquad (1)$$

Where $t_c$ is the current playhead position in the video, $S_W$ is the size of smoothing window, while $L^j$ is the quality level achieved at time period $j$. The average run metrics can be explained using Figure -9. In this figure, the average run for the time period $t_6$ can be calculated as

$$\bar{L} = \frac{L^{t_0}+L^{t_1}+L^{t_2}+L^{t_3}+L^{t_4}+L^{t_5}}{6} = \frac{3+1+1+1+2+2}{6} = 1.66.$$

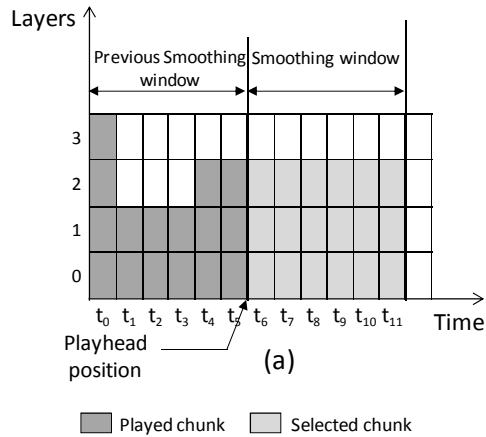

Figure -9: Illustration of Average run metrics

We use the average quality $\bar{L}$ in our algorithm for selecting the quality level for next time period, as shown in Figure-10. Our algorithm relies on the mean deviation $\alpha^t$ to decide the quality level $L^t$ at time $t$. We define $\alpha^t$ as:

$$\alpha^t = |L^{t-1} - \bar{L}|, \qquad (2)$$

Where $L^{t-1}$ is the tradeoff quality level at the previous time $t-1$. Let $B^t$ denote available download bandwidth of the receiver peer at time $t$, and $\widehat{B^t}$ the estimated bandwidth at time $t$. Further, let $\widehat{L^t}$ be the maximum quality that could be sustained with a bandwidth of $\widehat{B^t}$, and let $l_r^t$ represent the quality level which can be achieved using the remaining bandwidth in the period $t$. Then, the following algorithm can be used to calculate the tradeoff quality $L^t$ at the current time $t$.

***Smooth layered stream Procedure***

**if** $\left(\widehat{B^t} \geq B^{t-1}\right)$

$\qquad L^t = \min(\widehat{L^t}, L^{t-1} + \alpha^t)$

**else**

$\qquad L^t = \min(\widehat{L^t} + L_r^t,\ L^{t-1})$

**end if**

Figure-10: Smooth layered stream procedure

The algorithm for amplitude reduction takes two distinct cases into account. Whenever there is an expected increase in available bandwidth, the algorithm provides an increase in the quality level without violating the available bandwidth constraint at the receiving peer. The remaining available bandwidth is used to acquire the chunks in the prefetching buffer. In case of an expected decrease in available bandwidth, the algorithm reduces the amplitude by utilizing the already available chunks. Thus, the algorithm for amplitude reduction focuses on stepwise increase in order to reduce the jump when bandwidth decreases.

ii. *Frequency Reduction*

The objective of a frequency reduction mechanism is to reduce the number of quality level changes in the layered stream that can occur as a result of varying available bandwidth at the receiver peer. In this case, our objective will be to maintain the same quality level within a particular smoothing window. The frequency reduction mechanism is initialized by selecting the lowest quality for the first smoothing window. Then, the remaining available bandwidth is utilized to acquire the future chunks (these are of course stored in the prefetching buffer). The frequency reduction mechanism can be explained using Figure 11.

In Figure 11(a), the algorithm is initialized with the lowest quality level for the first smoothing window ($L^t = 0\ \forall t \in [t_0, t_5]$). The remaining available bandwidth can then be used to prefetch chunks which may be useful during the second smoothing window ($t \in [t_6, t_{11}]$). At the end of first smoothing window, the highest selected quality level in the prefetching buffer is 1. If the $\widehat{B^t}$ as estimated from data in $[t_0, t_5]$ is sufficient to acquire all the missing chunks for layer 1, then the quality level will be increased for the next smoothing window and $L^t = 1 \ \forall t \in [t_6, t_{11}]$, as shown in Figure 11(b). For frequency reduction the value of $\widehat{B^t}$ is calculated using the Extended Weighted Moving Average (EWMA) of the available bandwidth in the last smoothing window. The objective of using EWMA is to predict the bandwidth for the next smoothing window (instead of single time window as in amplitude reduction). On the other hand, if the chunks available in the prefetching buffer are insufficient to sustain the current quality level, as it happens for the third smoothing window in Figure 11(b), $L^t$ will be correspondingly reduced. In this example, we have that $L^t = 0 \ \forall t \in [t_{12}, t_{17}]$.

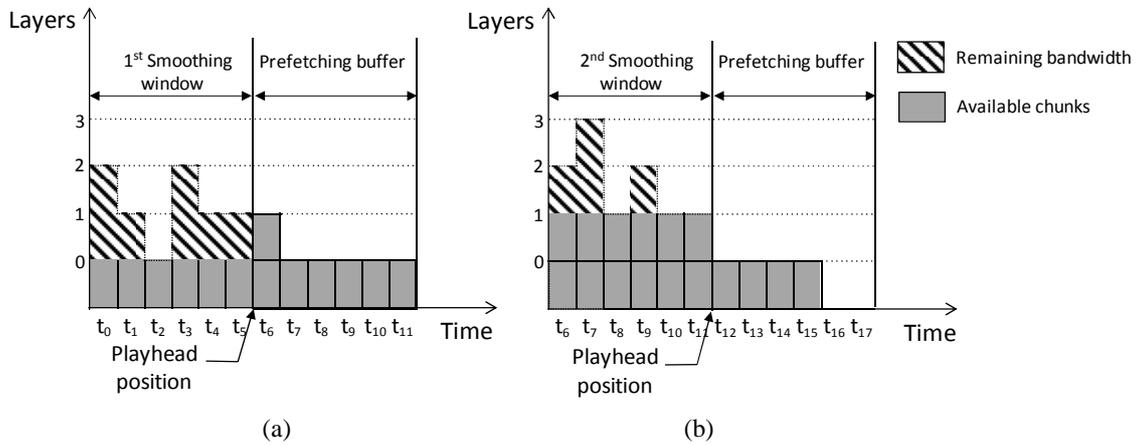

Figure 11: Frequency reduction mechanism

The main idea behind the frequency reduction algorithm is to initiate the quality level at 0 (equivalent to the base layer only) in the first smoothing window, and then use the remaining bandwidth to decide the quality level of the next smoothing window in a conservative manner, acquiring all lower layer chunks before than the higher layer ones. Thus the frequency smoothing mechanism ensures a constant quality level for each smoothing window. The bandwidth change in current smoothing window will ultimately affects the quality level in next smoothing window.

iii. *Hybrid approach*

The amplitude and frequency reduction mechanisms described earlier focus only on a single aspect of smoothing. We now propose a hybrid approach that provides the benefits of both the amplitude and frequency reduction mechanisms described earlier. To describe this approach, we label each smoothing window with an integer *k*, so that $L(k)$ denotes the constant stream quality selected for smoothing window *k*. Our hybrid approach is based on the frequency reduction mechanism presented earlier but also implementing amplitude reduction between successive smoothing windows. We define $\alpha(k)$ as the quality level difference between the current and the previous smoothing windows, as (see Figure 12)

$$\alpha(k) = |L(k) - L(k-1)|. \qquad (3)$$

Then, the quality level for the *k + 1* smoothing window can be calculated with

$$L(k+1) = L_r(k+1) + \beta(k), \qquad (4)$$

Where $L_r(k+1)$ is the quality level that is available at the beginning of smoothing window $k+1$ due to prefetching, and $\beta$ is chosen so that $|L(k+1) - L(k)| \leq \alpha$ with respect to the maximum quality level allowed by $\widehat{B^t}$ at the beginning of smoothing window $k+1$. The appropriate value of $\beta$ limits the change in the quality level among two successive smoothing windows.

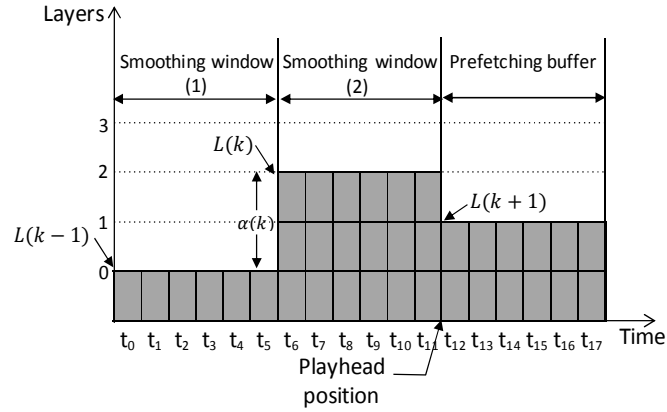

Figure 12: Hybrid approach

*B. The Scheduling Function*

The main goal of the scheduling function is to efficiently request the missing chunks in the exchanging window of the receiver peer. This can be achieved by requesting the higher priority chunks before the lower priority chunks while at the same time taking full advantage of the available network capacity. Since, this scheme will closely depend on the definitions of these priorities, we now explain how they are calculated.

Intuitively, it seems clear that since chunks are useless if they are not decoded by their playback deadline, the priority of each chunk should be closely related to how close they are to it. Another issue to consider is the dependency between layers; a higher layer chunk received without its corresponding lower layer chunks will not be decoded. To factor these two variables into our priority model, we will define two functions. The first one, the *emergency priority* $P_E$, is a function of how close a chunk is to its playback deadline; the second one, the *layer priority* $P_L$, is a function of how many underlying layers are necessary to decode a particular chunk. Using these two functions, we can define our priority function $P_{ij}$ as:

$$P_{ij} = P_E(T_i - D_j^i) + \theta P_L(L_j) \quad (5)$$

where $T_i$ denotes the current time in the peer $i$, $D_j^i$ denotes the playback deadline of chunk $j$ in peer $i$, $L_j$ denotes the stream layer to which chunk $j$ belongs, and $\theta$ is a parameter that can be adjusted for different layers prioritization strategies. Hence, $P_E(T_i - D_j^i)$ evaluates $P_E$ at a time interval equal to the remaining time that chunk $j$ has until its playback deadline at peer $i$, and $P_L(L_j)$ evaluates $P_L$ at an integer proportional to the number of underlying layers needed to decode chunk $j$.

Different values of $\theta$ can be used to implement different protocol behaviors. A small $\theta$ leads to the prioritization scheme represented in Figure 13(a), also called conservative chunk scheduling, where the receiver always requests chunks of lower layers first; a large $\theta$ leads to the aggressive chunk scheduling scheme represented in Figure 13(b), where chunks are requested on the basis of their timestamp only. Intermediate values of $\theta$ lead to tradeoffs between these two extremes; a particular example of this is shown in Figure 13(c).

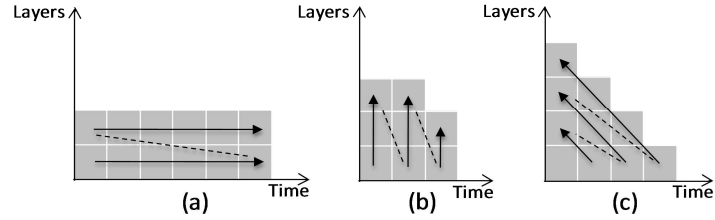
Figure 13: Scheduling strategies in case of layered streaming

We continue the presentation of our scheduling algorithm by defining $R_{ij}^k$, a Boolean variable that indicates whether the peer $i$ requests the chunk $j$ from the neighbor $k$:

$$R_{ij}^k = \begin{cases} 1, & \text{if peer } i \text{ requests chunk } j \text{ from neighbor } k, \\ 0, & \text{otherwise.} \end{cases}$$

We now present the core of our chunk scheduling heuristic. Using $P_{ij}$ as defined in (5), we propose the *aggregate priority* $\Pi_i$ of peer $i$ as a figure of merit for our scheduling algorithm:

$$\Pi_i = \sum_{\substack{j \in M_i \\ k \in N_i}} P_{ij} R_{ij}^k, \quad (6)$$

where $M_i$ denotes the set of chunks that peer $i$ requires from the overlay, and $N_i$ denotes the overlay neighbors of peer $i$. Using this figure of merit, our scheduling problem can be formulated for each peer $i$ as:

Maximize: $\Pi_i$

$$\text{Subject to:} \quad \sum_{j \in M_i} R_{ij}^k \leq C_i^k \qquad (7)$$

$$\sum_{k \in N_i} R_{ij}^k \leq 1 \qquad (8)$$

Where $C_i^k$ denotes the download capacity of the link between the receiving peer *i* and its neighbor *k*. Our scheduling mechanism will therefore maximize a figure of merit that trades off chunk urgency with stream quality, subject to a constraint on total link capacity (7). Equation (8) ensures that a receiver peer does not request a chunk *j* from more than one neighbor. That means a chunk is not requested more than once. Usually, each chunk is requested from a single neighbor, unless it is not available in the neighborhood. In this case it can't be requested.

This optimization problem can be naturally transformed into an Assignment Problem (AP) [23] where a set of missed chunks b∈ $M_i$ in peer i are to be assigned to a set $N_i$ of its neighbors. The assignment itself is captured with $R_{ij}^k$, the cost function for this assignment is $-\Pi_i$, and the feasibility conditions of the problem are (7) and (8). Therefore, in this case the set of chunks can be understood as a set of tasks which should be assigned to a set of agents (neighbors peers) while optimizing the overall cost, which refers to the priority sum of the chunks. In its original version, the AP involves assigning each task to a different agent, with each agent being assigned at most one task, i.e. *one-to-one assignment*. Since we want to assign one or more chunks to each neighbor, we will use an alternative formulation, the Generalized Assignment Problem (GAP) [24], that considers *one-to-many assignment* (multiple tasks can be assigned to the same agent). We therefore model the scheduling problem in layered streaming as a GAP, scheduling *m* chunks to *n* nodes (*m* ≥ *n*). This can be represented by the assignment matrix shown in Figure 14.

| Chunk \ Node | 1 | 2 | … | m-1 | m |
|---|---|---|---|---|---|
| 1 | $P_{i1}$ | $P_{i2}$ | … | $P_{i(m-1)}$ | $P_{im}$ |
| 2 | $P_{i1}$ | $P_{i2}$ | … | $P_{i(m-1)}$ | $P_{im}$ |
| … | … | … | … | … | … |
| n-1 | $P_{i1}$ | $P_{i2}$ | … | $P_{i(m-1)}$ | $P_{im}$ |
| n | $P_{i1}$ | $P_{i2}$ | … | $P_{i(m-1)}$ | $P_{im}$ |

Peers' reliability

Figure 14: Assignment matrix-GAP

The GAP is known to be NP-hard problem [24]. In the following section, we propose a novel heuristic to approximate its solution, and use it to perform chunk scheduling in Pull-based P2P streaming systems.

*Algorithm*

In order to construct a solution for the scheduling problem in layered streaming, modeled as GAP, we consider an arbitrary algorithm (let say algorithm A) to provide solutions (approximate or otherwise) for small versions of the knapsack problem (e.g. the Harmony-search algorithm [25]). As a first step, we reorganize the rows of the assignment matrix based on neighbors' reliability (Figure 14) in order to assign chunks to the more reliable nodes first. Then, we perform the recursive procedure shown below. Of course, *j* is initialized to 0 as part of the algorithm setup. In this algorithm, $N_i$ denotes the list of peer *i's* neighbors.

---

*Assignment_Matrix_Line_Processing*

1. Run the Algorithm A on the row *j* with respect to the capacity $C_i^k$ of node *k* and chunk size *r*. This will give the set of chunks from the most reliable peer that has not been considered yet, and which maximize aggregate priority. Let $S_j$ be this solution, i.e. set of selected chunks returned.

2. **if** $j < |N_i|$ *(Termination condition)*

    - j = j + 1. *(Increment the indicator variable j, so that recursive calls to this function will consider the next row of A).*
    
    - Perform *Assignment_Matrix_Line_Processing*(*j*) and let $S'$ be the returned chunks list. Return $S_j \cup S'$

   **else**

   Return $S_j$

---

Figure 15: Assignment matrix line processing algorithm

*C. Processing Overhead*

The availability of high speed computing allows processing the huge dataset for ARIMA to build a statistical model [26], however, in this work we only consider the traffic samples for last smoothing window thus reducing the overhead of processing large amount of data. The ARIMA model is lightweight in memory and calculation cost and even used in sensor devices. The model buildup process may take relatively high memory and computational overhead, but the process is done by the receiver peer, which usually has high computational and storage capability [27].

In addition, our scheduling algorithm is based on a powerful knapsack algorithm, mainly the Harmony search algorithm. The results obtained using the HS algorithm may yield better solutions than those obtained using current algorithms [28], such as conventional mathematical optimization algorithms or genetic algorithm based approaches. The study performed in [28] suggests that the HS algorithm is potentially a powerful search and optimization technique for solving complex engineering optimization problems.

Finally, the amount of data processed in each scheduling period is very low. Indeed, the exchanging window size is very low, and the maximum number of neighbors considered in the simulation is not more than 30 neighbors, consequently the scheduling matrix size is small, and can be proceeded in very short time.

## V. PERFORMANCE EVALUATION

In this section, we present the performance evaluation of our proposed playout smoothing mechanism for layered streaming (referred as *SmoSched* in the simulations). We first need to define the relevant metrics that reflect the key features of perceived video quality and the effective utilization of the available network capacity.

We focus mainly on two important points: first, we evaluate the performance of smoothness for our proposed mechanism, mainly in terms of layers changes and stalling events. Secondly, we evaluate the throughput of the system in terms of bandwidth utilization, useless chunks and delivery ratio.

### A. Metrics

We consider the following metrics for the performance evaluation of our proposed mechanism.

***Number of layer changes during video playback:*** We measure the average number of changes in the quality level during the video playback.

***Stalling Events:*** We are interested in the average number and duration of stalling events that occur during video playback due to the unavailability of the content. The shorter the stalling event, the better is the session quality.

***Bandwidth utilization:*** The effective utilization of the available bandwidth is an important consideration in P2P streaming. It measures the effective bandwidth used over the total available bandwidth in the network.

***Useless chunks ratio:*** It represent the number of chunks that arrive at each peer after their playback deadline over the total number of encoded chunks.

***Delivery ratio at layer l:*** It is defined as the average delivery ratio at layer *l* among all the peers that can play layer *l*. A chunk of layer *l* is considered as properly received if and only if all the related chunks of lowers layers to *l* are already received before their playback deadline.

## B. Evaluation

Here we present our simulative evaluation for the proposed mechanism. The goal of this study is to observe the behavior of the metrics discussed. To effectively evaluate the performance of each component and compare it with different existing techniques, we perform the evaluation into two steps.

### a) Scenario 1

In this scenario, we evaluate the metrics which had considerable importance in observing the behavior of smooth layered stream. This includes layer changes and stalling events. We compare the proposed mechanism with LayerP2P [16] discussed earlier. We used the BRITE universal topology generator [28] in the top-down hierarchical mode to map the physical network. The network topology consists of autonomous system (AS) and fixed number of routers. All AS are assumed to be in the Transit-Stub manner. Each topology consists of 8 autonomous systems each of which has 625 routers. This gives us about 20000 links in the topology. The delay on inter-transit domains and intra-transit domains are assumed to be 90 ms and 40 ms respectively, while delay on stub-transit is assumed to be 30 ms and intra-stub transit links are randomly chosen between 5ms and 30ms. The incoming bandwidth of peers varies between 512 kbps to 2Mbps and is uniformly distributed throughout the network. The video is composed of 4 layers streamed at 1Mbps. We introduced sudden bandwidth change in the network by varying peer's capacity. This allows us to observe the effectiveness of smoothing mechanism.

**Results and Discussion**

Figure 16 shows the number of layer changes at different intervals for both mechanisms in an overlay composed of 300 peers. The objective was to minimize the number of layer changes that occurs due to changing network condition. The simulations are performed for different values of $\beta$ while keeping into consideration bandwidth constraint $\widehat{B^t}$. The interval on x-axis represents the time after which the bandwidth change occurs. It is observed that layer changes decreases by increasing the interval. Thus, it is found that bandwidth change has an obvious effect on the variation of the quality. Our proposed mechanism has very fewer layer changes due to the smoothing mechanism which adjusts the quality level for each smoothing window. As a result, the frequent changes in the quality level are minimized. Comparatively, LayerP2P system has higher number of quality changes due to the absence of an efficient smoothing mechanism.

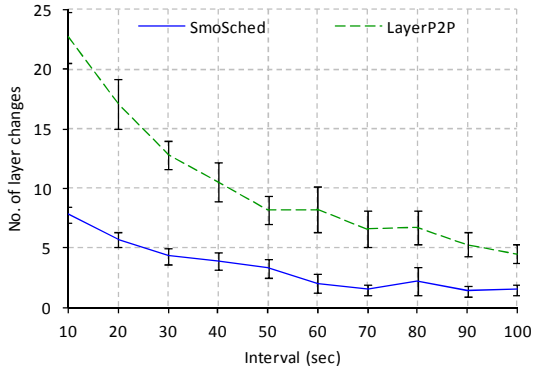
Figure 16: No. of Layer Changes

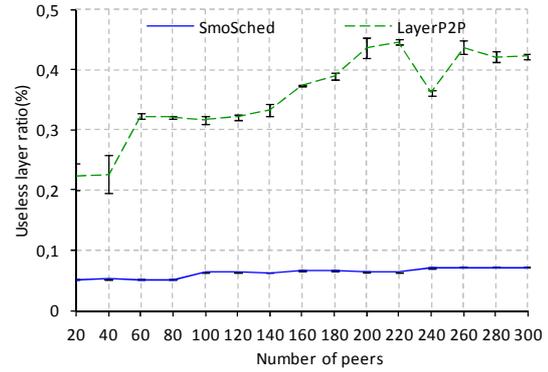
Figure 17: Useless layer ratio

Figure 17 shows the comparison for useless layer ratio at different network size configurations. The higher layers which are selected and are not decoded due to the sudden decrease in bandwidth results in useless layers. Our proposed mechanism ensures the moderate selection of higher layer when a bandwidth increase occurs. Instead of allocating all the available bandwidth for acquiring maximum quality, it focuses on allocating the available bandwidth for the prefetching buffer. The moderate layer selection allows the scheduling mechanism to acquire the chunks of selected layer without violating the bandwidth constraint. As a result the selected layers are effectively decoded. The LayerP2P mechanism has higher useless layer ratio because there is aggressive selection of layers without taking into consideration the bandwidth constraint.

Figure 18 shows the average number of stalling events for both mechanisms. The stalling of video stream occurs due to the unavailability of selected layer chunks at that particular time period. The maximum number of stalling event in our case was not more than 6. However, there is a huge variation of stalling events in case of LayerP2P. Moreover the average duration of stalling event is lower in our proposed mechanism as shown in Figure 19.

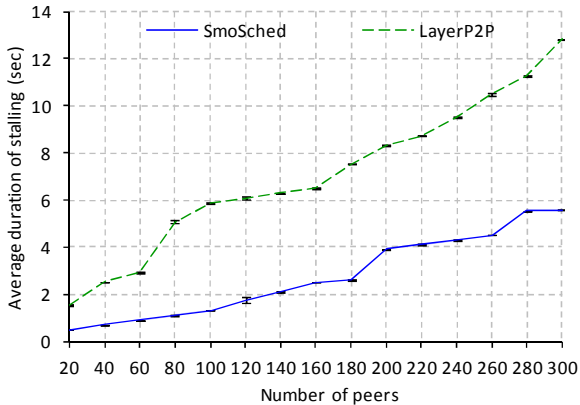
Figure 18: Average duration of stalling events

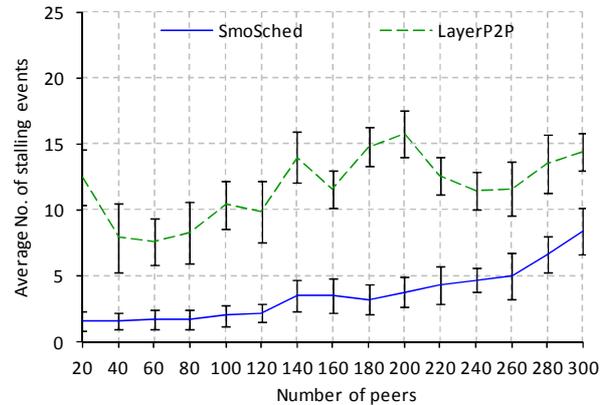
Figure 19: Average No. of stalling events

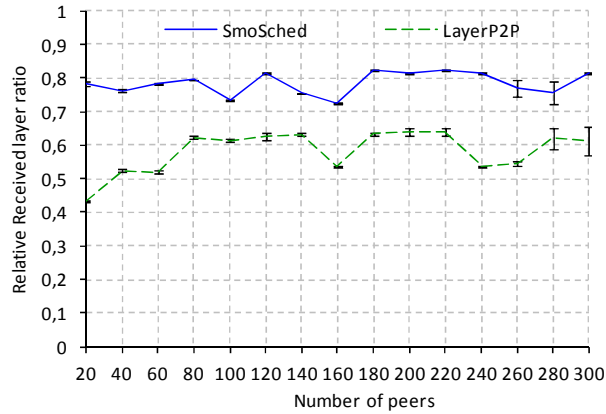

Figure 20: Relative Received layer ratio

Figure 20 shows the comparison of relative received layers for both mechanisms. The relative received layer represents the layers which are delivered completely. It is observed that more than 85% requested layers are delivered completely in proposed mechanism. This shows the effectiveness of proposed mechanism. On the other hand LayerP2P has lower relative delivery ratio due to unavailability of lower layers chunks which ultimately affects the higher layers.

In order to evaluate the other smoothing strategies discussed earlier, namely amplitude reduction and frequency reduction, we consider a video stream composed of 8 layers (each layer is streamed at 100 kbps) under bandwidth variation (from 100 kbps to 900 kbps). The duration of the video is 400 seconds and the smoothing window size is 15 seconds.

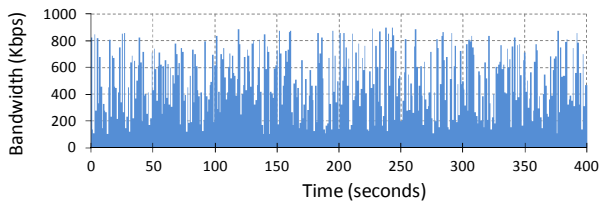

Figure 21: bandwidth variation

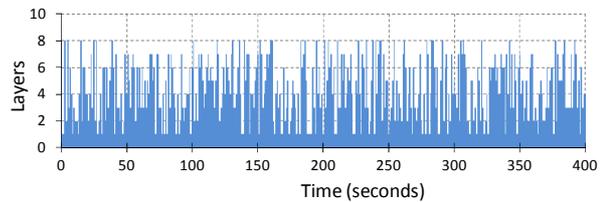

Figure 22: Row stream

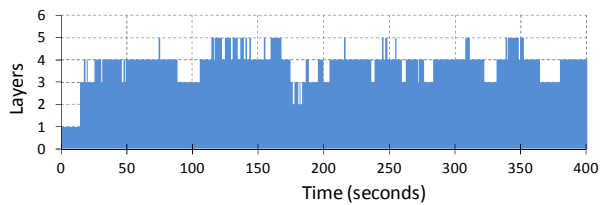

Figure 23: Amplitude reduction

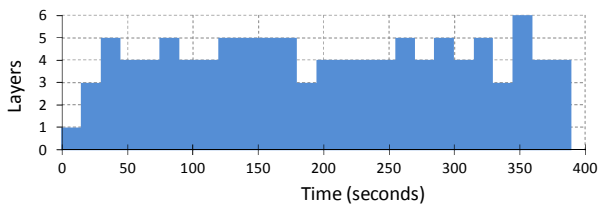

Figure 24: Frequency reduction

In Figure 21 we plotted the aggregated download bandwidth in the receiver peer. Figure 22 represents the layers changes without any smoothing action on the stream. In this figure we note that the quality level fluctuates with the fluctuation of the aggregated upload bandwidth of the peer. In Figure 23 we apply the amplitude reduction algorithm on the row stream. In this case we note that the fluctuation of the quality is drastically reduced, and the user can enjoy a relative stable video quality comparing to row stream. Indeed, this figure shows that the quality level change does not exceed 1 level in most cases. However, they could be a fluctuation of the quality within a short time period (from second 100 to 150 for example). In Figure 24, we apply the frequency reduction algorithm on the same row stream. We note that the quality level is stable at least for a smoothing window, however we note some big jump in the quality level in some cases (jump of 2 layers from 170s to 180s for example). This is why we have proposed a hybrid smoothing solution which combines the benefits of the amplitude reduction and the frequency reduction algorithms.

*b) Scenario 2*

In this scenario, we evaluate the efficient delivery of the content considering the availability of the content among the neighbors and their corresponding bandwidth capacity. For that purpose, we focus on the different metrics including bandwidth utilization, useless chunks ratio and delivery ratio at different layers.

We perform extensive simulations using an overlay in which each peer has varying number of neighbors. We compare the performance of our algorithm with three classic scheduling methods described earlier, namely Random strategy (RND), Local Rarest First (LRF) and Round Robin (RR). We consider three categories of peers: 40% of users with 512Kbps, 30% with 1Mbps and 30% with 2Mbps, and for all users, the upload bandwidth capacity is half of the download bandwidth. We set the emergency priority defined in (5) as $P_E(T_i - D_j^i) = 10^{(T_i - D_j^i)}$ and we set the layer priority as $P_L(L_j) = 10^{(L-L_j)}$ in order to ensure that the lower layers have much larger priority than the upper layers. For the four methods, we adopt the conservative approach (Figure 13) described in section IV. That's why we set the parameter $\theta$ to a very low value $\theta = 10^{-L}$.

**Results and Discussion**

In Figure 25, we study the performance of our mechanism in terms of bandwidth utilization and we examine the effect of the neighbor density. It is observed that our proposed mechanism outperforms the three other scheduling schemes and ensures more than 90% of bandwidth utilization, while the LRF scheme allows the maximum bandwidth utilization of 88%. The RND method shows the worst result (up to 71%). On the other hand, we note that the bandwidth utilization for all the schemes increases with the increase in the number of neighbors (for maximum 15 neighbors). This is due to the increase of the chunks availability in the neighborhood. The stable behavior is observed when the number of neighbors for a peer is greater than 15. We conclude that, for the considered overlay, the average of 15 neighbors per peer is enough to ensure the chunks availability.

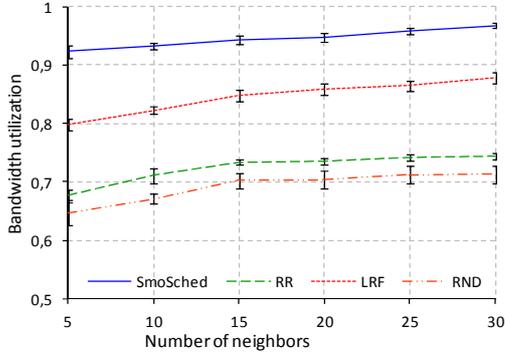

Figure 25: Bandwidth utilization Vs. Neighbor density

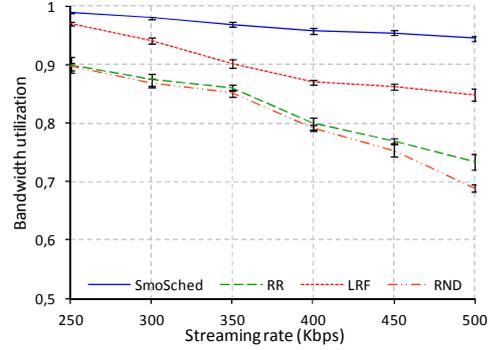

Figure 26: Bandwidth utilization Vs. Streaming rate

Moreover, we are interested on the impact of the streaming rate on the bandwidth utilization. The results are presented in Figure 26. We note that the general trend is the decreases in bandwidth utilization with the increasing streaming rate. This can be explained by the fact that when the streaming rate is high, a missed chunk has less chance to be rescheduled before its playback deadline. Nevertheless, our proposed mechanism outperform the three others mechanisms and allows a gain up to 25%. This is due to the fact that our mechanism tries to fully take advantage of the available bandwidth in the neighbor, using the assignment mechanism.

In Figure 27 and Figure 28, we evaluate the useless chunks ratio according to the neighbor's density and the streaming rate respectively. It is observed in Figure 27 that the useless chunks ratio decreases with the increasing number of neighbors. This decrease of useless chunk ratio is due to the higher probability of getting the requested chunks from the large pool of neighboring peers. Moreover, our mechanism tries to find the good tradeoff between the chunks availability in order to request the right chunk from the right neighbor. That's why the useless chunks ratio is low in proposed mechanism compared to the others systems.

In Figure 28, we evaluate the useless chunks ratio with the varying streaming rate. It is found that the useless chunks increase with the increase of the streaming rate. However, the proposed mechanism still outperforms the existing systems.

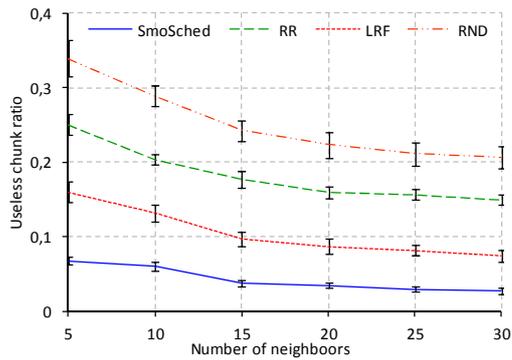
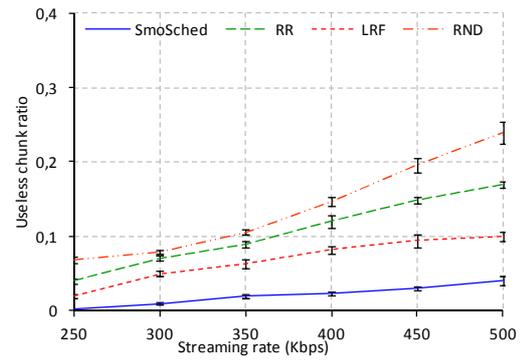

Figure 27: Useless chunks ration Vs. Neighbor density

Figure 28: Useless chunks ration Vs. Streaming rate

Figure 29 describes the delivery ratio at each layer. We encode the video into 12 layers and set the rate of each layer at 100 Kbps. We note that our proposed mechanism is fairly good. In lower layers, delivery ratio is approximately 1 and in higher layers it is also above 0.9. The RR has much more better delivery ratio at lower layers than higher layers but, the delivery ratio for all layers is not as good as the proposed mechanism. We note that the LRF strategy has even higher delivery ratio than the RR strategy. Finally, the random strategy has the poorest performance. As shown in Figure 29, our proposed mechanism outperforms other strategies with a gain of 10%-50% in most layers.

In order to show the importance of varying number of layers, we encode the video into 6 layers. In Figure 30, we note that the delivery ratio of each layer is nearly similar to that in 12 layers encoding scenario. Our mechanism is still the best among all the three others methods. However, we note that the delivery ratio of all the methods is little higher than in the case of 12 layers. This is due to the fact that encoding the video into six layers allow peers to allocate all their bandwidth to lower layers, however in the second case, some bandwidth will be dedicated to the higher layers (higher than 6).

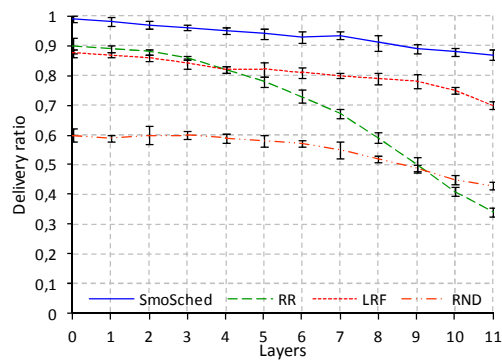
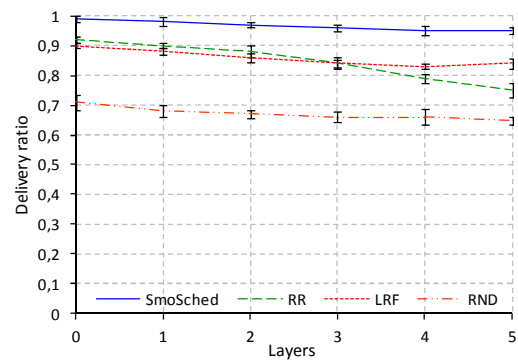

Figure 29: Delivery ratio (12 layers)

Figure 30: Delivery ratio (6 layers)

## VI. CONCLUSION

In this paper, we propose a playout smoothing mechanism for layered streaming in P2P network that selects appropriate stream layers while minimizing the number of layer changes and achieve high delivery ratio. Firstly, we proposed the mechanism for amplitude and frequency reduction in layered streaming. We then combine the benefits of both mechanisms in hybrid approach. The proposed mechanism first selects the appropriate layers and then schedules those layers to the appropriate neighbors thus effectively utilizing the available capacity of the network.

To study the effectiveness of the proposed mechanism, we performed the simulation and compared the proposed mechanism with different existing systems. We studied different metrics that are essential in determining the performance of our proposed playout smoothing mechanism. The results demonstrate the optimality and the effectiveness of the solution.

The main limitation of *SmoSched* is the amount of delay introduced by the frequency reduction component, in particular, the smoothing window size parameter. As future work, we plan to study techniques to achieve a trade-off between the smoothing quality and the liveness of the stream by acting on the smoothing window size.